\documentclass[twocolumn,showpacs,amsmath,amssymb,showkeys,floatfix,prl]{revtex4}
\usepackage{amsfonts,amsmath}
\usepackage{graphicx}
\usepackage{dcolumn}
\usepackage{bm}

\begin{document}

\title{Neutrino oscillations: Measuring $\theta_{13}$ including its sign}

\author{D. C. Latimer and D. J. Ernst}

\affiliation{Department of Physics and Astronomy, Vanderbilt University, 
Nashville, Tennessee 37235, USA}
\date{\today}

\begin{abstract}
In neutrino phenomenology, terms in the oscillation
probabilities linear 
in $\sin \theta_{13}$ lead naturally to the question
``How can one measure $\theta_{13}$ including its sign?'' 
Here we demonstrate analytically and with a simulation of neutrino data 
that 
${\mathcal P}_{e\mu}$ and ${\mathcal P}_{\mu\mu}$ at $L/E = 2\pi/\Delta_{21}$ 
exhibit significant linear dependence on
$\theta_{13}$ in the limit of vacuum oscillations.  Measurements at this
particular value of $L/E$ can thus determine not only $\theta_{13}$ 
but also its sign, if CP violation is small. 
\end{abstract}



\maketitle
\section{Introduction}

To incorporate neutrino oscillations the standard model is conventionally  extended by adding a mass term and a mixing 
matrix. This theory of three-flavor neutrino oscillations has been successful in accommodating the results of
neutrino oscillation experiments, save LSND \cite{LSND}.  The theory contains six independent parameters:  three mixing
angles $\theta_{jk}$, two independent mass-squared differences $\Delta_{jk}:= m_j^2 - m_k^2$, and one Dirac CP phase $\delta$.
Ref.\ \cite{bargerrev}, for example, provides a summary of the current knowledge of the values of these parameters 
and describes future experiments.
The long baseline (LBL) K2K experiment \cite{k2k} is in the process of confirming the results of the Super-K atmospheric
experiment \cite{superk} by measuring the parameters $\Delta_{32}$ and $\theta_{23}$, independent of atmospheric neutrino flux
models.  Future LBL experiments, MINOS \cite{minos}, OPERA \cite{opera}, and ICARUS \cite{icarus}, will improve upon the
bounds for these parameters.  Additionally, a global analysis of these LBL experiments could
provide a lower bound on the magnitude of $\theta_{13}$ \cite{lblanalysis}. 
Future LBL experiments might also resolve the question of mass hierarchy and the level, if any,
of CP violation in the neutrino sector \cite{eightfold}.  

We here examine the related question of how to best measure $\theta_{13}$, including its sign.
In Ref.~\cite{angles}, we have shown that, even in the presence of matter effects \cite{msw}, neutrino oscillations can be uniquely
and completely parameterized with the following bounds on the angles: the CP phase $\delta$ lies in the range $[0, \pi)$;
$\theta_{13}$ lies in the range $[-\pi/2, \pi/2]$; and the remaining mixing angles lie within the first
quadrant. This choice of bounds has two advantages. 
First, present experiments limit $\theta_{13}$ to a small asymmetric region
about zero \cite{nolsnd3,us}. Other choices would break this region into two disconnected regions. 
Secondly, the CP violating
phase is restricted to the first two quadrants; this range is sufficient to characterize all CP violating effects. 
Terms proportional to $\cos\delta$ are thus able to uniquely determine its value, assuming knowledge of all the other parameters.

In the next section, we analytically examine the terms of the neutrino oscillation probability formulae that are first order (linear)
in $\theta_{13}$. These terms are proportional to either $\sin\delta$ or $\cos\delta$, as indicated in 
Refs.~\cite{bargerrev,smirnov1}.  It has been suggested \cite{smirnov1,eexcess} that the presence of such terms, in part, 
can explain the excess of electron-like events in the Super-K atmospheric experiment \cite{superk}.  In this work, we find that experiments
which lie in the oscillatory region for the small (solar) mass-squared difference, an $L/E$ on the order of 10$^{4}$ m/MeV, are
sensitive to the linear $\theta_{13}$ terms. We further find that by judicious choice of the value of $L/E$ the effects of the
CP violating phase $\delta$ can be suppressed if $\delta$ is near zero or $\pi$, while at the same time the effect of the
linear term in $\theta_{13}$ is maximized. In the subsequent section, we utilize a simulation \cite{us} of 
the existing neutrino oscillation data to
further examine the ability of new data to determine $\theta_{13}$, including its sign. We assume that either CP is conserved or that
our choice of the value $L/E$ has provided sufficient suppression of the CP violating terms. 
In the final section we summarize our conclusions and provide some 
thoughts on needed future theoretical work.

\section{Formal analysis}

In this section we provide explicit analytic expressions for three neutrino oscillations valid
for the incoherent limit of the atmospheric mass-squared difference. 
We confine our discussion to vacuum oscillations.  
For LBL experiments through the earth, we indicate which values of $L$ and $E$ yield the cleanest measurement
of $\theta_{13}$ by avoiding significant contributions from matter effects \cite{msw}. Additionally, we indicate qualitatively 
the consequences of straying outside these energies and baselines.
We use these analytic expressions to examine where the effects of the linear terms
can best be seen. As the magnitude of $\theta_{13}$ and the mass-scale ratio 
$\alpha := \vert \Delta_{21} \vert / \vert \Delta_{32}\vert$ are known to be small, one may expand the oscillation probability formulae 
about these parameters.   In these perturbations (cf.\ \cite{expans}), terms which are linear in $\theta_{13}$
are suppressed by a factor of $\alpha \sim 0.03$.  
From this, one might conclude that effects relevant to the sign 
of $\theta_{13}$ are
forever relegated to the realm of the unobservable. 
Here, this is not the case as we look beyond the valid region of these expansions. 

We use the standard representation \cite{PDG} of the three-neutrino mixing
matrix with the notation  $c_{j k} = \cos{\theta_{jk}}$,
$s_{j k}=\sin{\theta_{j k}}$,  and $\delta$ is the CP violating phase. 
In a three-neutrino theory, the probability that a neutrino with relativistic energy $E$ and flavor $\alpha$ 
will be detected a distance $L$ away as a neutrino of flavor $\beta$
is given by
\begin{eqnarray}
\mathcal{P}_{\alpha \beta}(L/E)&=& \delta_{\alpha \beta}\nonumber\\
&&-4 \sum^3_{\genfrac{}{}{0pt}{}{j <
k}{j,k=1}} \mathrm{Re} (U_{\alpha j} U^*_{\alpha k} U_{\beta k} 
U^*_{\beta
j}) \sin^2 {\varphi_{jk}} \nonumber \\
&&+2 \sum^3_{\genfrac{}{}{0pt}{}{j < k}{j,k=1}} \mathrm{Im} (U_{\alpha
j} U^*_{\alpha k} U_{\beta k}
U^*_{\beta j}) \sin{2 \varphi_{jk}} \, , \nonumber \\  && \label{oscform}
\end{eqnarray}
where 
$\varphi_{jk} := \Delta_{jk} L/4E$ with $\Delta_{jk} := m_j^2 - m_k^2$. 

Examining the terms which are linear in $\sin \theta_{13}$ motivates us to consider the limit in which the oscillations due to the
mass-squared differences $\vert \Delta_{32} \vert \cong \vert \Delta_{31} \vert \sim 10^{-3} \mathrm{eV}^2$ are
incoherent while the oscillations due to $\vert \Delta_{21} \vert \sim 10^{-5} \mathrm{eV}^2$ are still relatively
coherent.  In this limit, we may take
\begin{equation}
\sin^2{\varphi_{23}} = \sin^2 {\varphi_{13}} = \frac{1}{2}, \quad \sin{2 \varphi_{23}} = \sin {2\varphi_{13}} = 0.
\end{equation}
The oscillation probabilities $\mathcal{P}_{e\mu}$, 
$\mathcal{P}_{\mu \mu}$, and $\mathcal{P}_{\mu \tau}$, in the limit of incoherent atmospheric oscillations, are then given by
\begin{eqnarray}
\mathcal{P}_{e \mu} &=& [\tfrac{1}{2} \sin{2\theta_{12}} \cos{2\theta_{12}} \sin{2\theta_{13}} c_{13} \sin{2\theta_{23}}
c_\delta \nonumber \\
&& + \sin^2{2 \theta_{12}} c_{13}^2(c_{23}^2-s_{13}^2 s_{23}^2)] \sin^2 {\varphi_{12}} \nonumber \\ 
&& + \tfrac{1}{2}\sin^2{2\theta_{13}} s_{23}^2 +2 J \sin{2 \varphi_{12}},  \label{pem}
\end{eqnarray}
\begin{eqnarray}
\mathcal{P}_{\mu \mu} &=& 1-[ \sin{2\theta_{12}} \cos{2\theta_{12}} \sin{2\theta_{13}} c_{13} \sin{2\theta_{23}}
s_{23}^2 c_\delta \nonumber \\
&&  + 2 \sin{2\theta_{12}} \cos{2\theta_{12}} s_{13} \sin{2\theta_{23}}
\cos{2\theta_{23}} c_\delta  \nonumber \\
&&  + (1-\sin^2 {2\theta_{12}} c_\delta^2) s_{13}^2 \sin^2 {2\theta_{23}}  \nonumber \\
&& +\sin^2{2\theta_{12}}(c_{23}^2 - s_{13}^2 s_{23}^2)^2] \sin^2 {\varphi_{12}}   \nonumber \\
&&   -2 c_{13}^2 s_{23}^2 (1-c_{13}^2 s_{23}^2), \label{pmm}
\end{eqnarray}
\begin{eqnarray}
\mathcal{P}_{\mu \tau} &=& [ \sin{2\theta_{12}} \cos{2\theta_{12}} s_{13}(1+s_{13}^2) \sin{2\theta_{23}}
\cos{2\theta_{23}} c_\delta \nonumber \\
&& (1-\sin^2{2\theta_{12}} c_\delta^2) s_{13}^2 \sin^2{2 \theta_{23}} \nonumber \\
&& -\tfrac{1}{4} \sin^2 {2\theta_{12}} (1+s_{13}^2)^2 \sin^2 {2\theta_{23}} \nonumber \\
&& + \sin^2 {2 \theta_{12}} s_{13}^2] \sin^2 \varphi_{12} \nonumber \\
&& + \tfrac{1}{2} \sin^2 {2 \theta_{23}} c_{13}^4 + 2J \sin{2 \varphi_{12}}, \label{pmt}
\end{eqnarray}
with 
\begin{equation} 
J= \tfrac{1}{8} \sin{2\theta_{12}} \sin{2\theta_{13}}c_{13} \sin{2\theta_{23}} s_\delta.
\end{equation}
As we are interested in the sign of $\theta_{13}$, 
we isolate those terms that are odd with
respect to $\theta_{13}$
\begin{eqnarray}
\lefteqn {\mathcal{P}_{e \mu}(\theta_{13}) - \mathcal{P}_{e \mu}(-\theta_{13}) = 4J \sin{2\varphi_{12}}+} \nonumber \\
&& \sin{2\theta_{12}} \cos{2\theta_{12}} \sin{2\theta_{13}} c_{13} \sin{2\theta_{23}}
c_\delta \sin^2 {\varphi_{12}}, \label{pemth13} \\
\lefteqn {\mathcal{P}_{\mu \mu}(\theta_{13}) - \mathcal{P}_{\mu \mu}(-\theta_{13}) =
-4(c_{23}^2-s_{13}^2 s_{23}^2) } \nonumber \\
&& \times \sin{2\theta_{12}} \cos{2\theta_{12}} s_{13} \sin{2\theta_{23}} 
c_\delta \sin^2 {\varphi_{12}}, \label{pmmth13} \\
\lefteqn {\mathcal{P}_{\mu \tau}(\theta_{13}) - \mathcal{P}_{\mu \tau}(-\theta_{13}) = 4J \sin{2\varphi_{12}}+2\sin{2\theta_{12}} } 
\nonumber \\
&& \times \cos{2\theta_{12}} s_{13}(1+s_{13}^2) \sin{2\theta_{23}} \cos{2\theta_{23}}
c_\delta \sin^2 {\varphi_{12}}. \nonumber \\ \label{pmtth13}
\end{eqnarray}
Note that the sign of $\theta_{13}$ exhibits the maximal effect whenever $\sin^2{\varphi_{12}}$ is
maximal for $c_\delta \sim 1$.  
This occurs whenever $\varphi_{12} = (2n+1) \pi/2$ or, in other terms, for $L/E = 2(2n+1) \pi/ \Delta_{12}$. 
These oscillations will be more coherent for the smaller values of $n$. 

This choice is fortuitous in that it also removes from 
Eqs.\ (\ref{pemth13}) and (\ref{pmtth13}) terms dependent on $s_\delta$ as $\sin{2 \varphi_{21}}=0$ 
whenever $\varphi_{21}$ is an odd-integer multiple of
$\pi/2$.  This removes the CP violating terms from consideration. The remaining terms are 
modulated by $c_\delta$.  Thus for $\delta$ near zero or $\pi$, we would have a clean measurement
of $\theta_{13}$.  Also, should CP violation be found to be maximal, then the terms involving $c_\delta$ vanish.

A consequence of removing the dependence on $J$ is that CP violating effects are suppressed. At these local values 
of $L/E$ we have $\mathcal{P}_{\alpha \beta} =
\mathcal{P}_{\overline{\alpha} \overline{\beta}}$, where $\overline{\alpha}$ indicates an 
antineutrino of flavor
$\alpha$; assuming CPT is invariant, this can be expressed as 
$\mathcal{P}_{\alpha \beta} =
\mathcal{P}_{\beta \alpha}$.  
This supports our previous statement that it is sufficient to only consider $\mathcal{P}_{e
\mu}$, $\mathcal{P}_{\mu \mu}$, and $\mathcal{P}_{\mu \tau}$ in regards to their dependence on the sign of
$\theta_{13}$.  These probabilities in addition to $\mathcal{P}_{ee}$, which is a function of
$\theta_{13}^{\, 2}$, will give us all the other oscillation probabilities at this value of $L/E$.  The remaining probabilities
are
\begin{eqnarray}
\mathcal{P}_{e \tau} = 1- \mathcal{P}_{e e} - \mathcal{P}_{e \mu}, \\
 \mathcal{P}_{\tau \tau} = 1- \mathcal{P}_{\mu \tau} - \mathcal{P}_{e \tau},
\end{eqnarray}
so that the dependence of $\mathcal{P}_{e \tau}$ on the sign of $\theta_{13}$ can be surmised from the statements 
made concerning $\mathcal{P}_{e \mu}$ and, likewise, the behavior of  $\mathcal{P}_{\tau \tau}$ can be surmised from 
$\mathcal{P}_{\mu \tau}$ and $\mathcal{P}_{e \mu}$. In what follows, we will assume that any
CP violation is small so that we may set the phase equal to 0.

\begin{figure}
\includegraphics[width=3in]{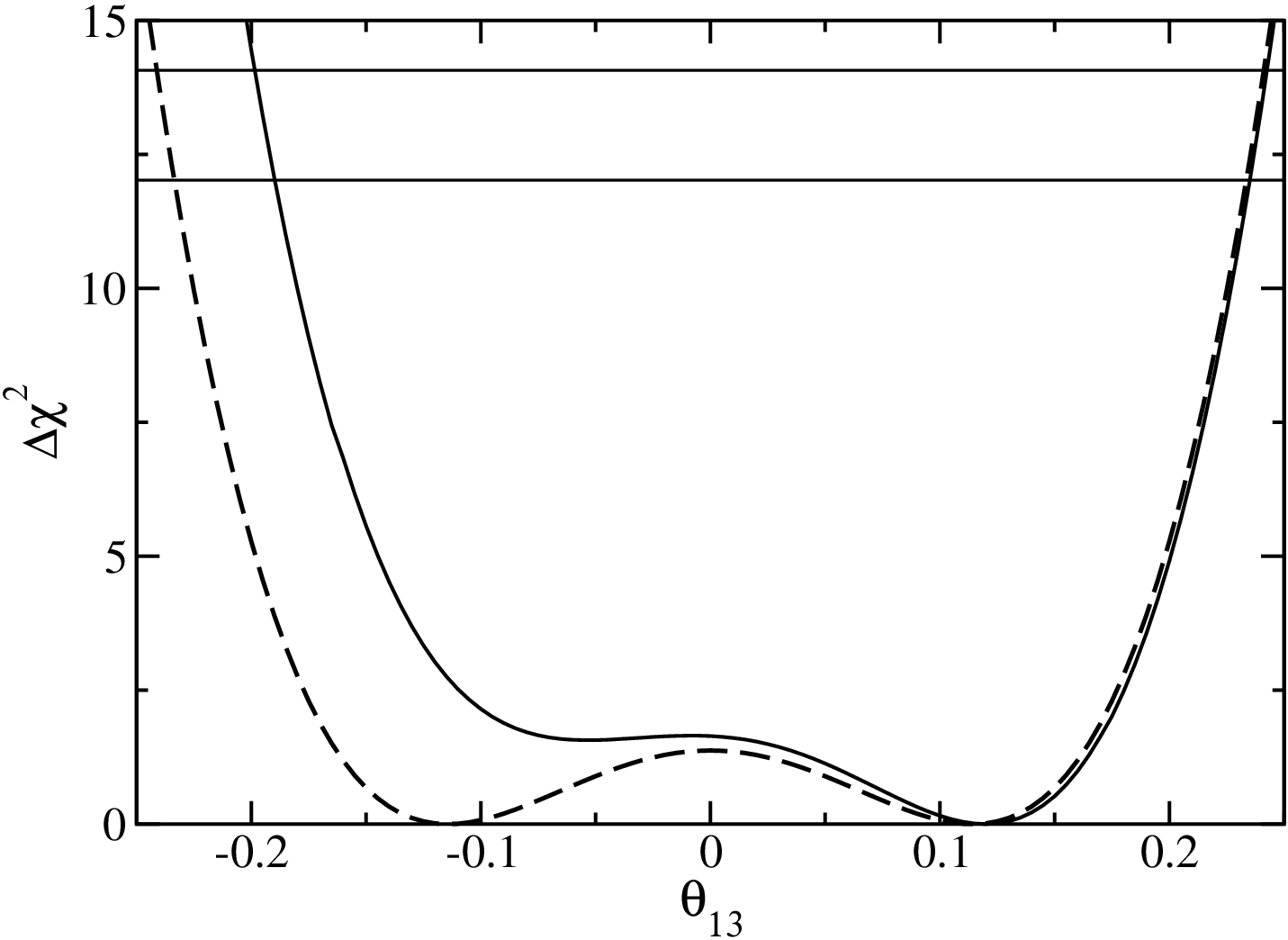}
\caption{The value of $\Delta\chi^2=\chi^2-\chi^2_{min}$ versus $\theta_{13}$ as extracted from the world's neutrino oscillation
data using the simulation of Ref.~\protect\cite{us}. All other parameters other than $\theta_{13}$ are varied. There are two minima, 
$\theta_{13}=0.11$ and $-0.04$. The dashed line represents the results when the ``one mass-squared dominance" 
approximation is utilized. The horizontal lines represent the 90\% and 95\% confidence levels.
\label{fig1}}
\end{figure}

We have previously performed \cite{nolsnd3} a simulation, assuming no 
CP violation, of the world's neutrino oscillation data. An analysis which includes more recent data \cite{us}, in
preparation, produces $\Delta\chi^2 := \chi^2 - \chi^2_{min}$ as pictured in Fig.~\ref{fig1}. Included in the analysis
are data for neutrinos from the sun \cite{solar}, from reactors \cite{reactors}, atmospheric neutrinos \cite{atmo},
and beam-stop neutrinos \cite{k2k}.
For one standard deviation,
the analysis bounds $\theta_{13}$ to lie within $[-0.17, 0.22]$ with two minima located at $\theta_{13} = 0.11$ and $-0.04$. For the absolute minimum
$\theta_{13}=0.11$, we find $\theta_{12}=0.48$, $\theta_{23}=0.80$, $\Delta_{21}=7.7\times 10^{-5}$ eV$^2$, and $\Delta_{32}=2.6\times
10^{-3}$ eV$^2$.

Since ${\mathcal P}_{ee}$ is a function of $\sin^2\theta_{13}$, the asymmetry seen in Fig.~\ref{fig1} must arise from the atmospheric and K2K data, which
involve ${\mathcal P}_{e\mu}$ and ${\mathcal P}_{\mu\mu}$. If we employ the ``one mass-squared dominance" approximation, as is often done, we find the
dashed curve presented in Fig.~\ref{fig1}. This approximation gives oscillation probabilities that are a function of $\sin^2\theta_{13}$.

In order to demonstrate the relative size of the effect of the sign of $\theta_{13}$, we choose some 
realistic values
for the mixing angles:  $\theta_{12}=0.56$ and $\theta_{23} = 0.78$. The first two peaks of $\sin^2{\varphi_{12}}$ occur 
around $L/E = 1.6 \times 10^4~
\mathrm{m/MeV}$ and $4.8 \times 10^4 ~\mathrm{m/MeV}$.  For such values of $L/E$, the oscillations due to $\Delta_{32}$
and $\Delta_{31}$ would be incoherent.   
It is clear from Eq.\ (\ref{pmtth13}) that the screening effect of
maximal mixing for $\theta_{23}$ results in independence of the sign of $\theta_{13}$ for $\mathcal{P}_{\mu \tau}$.  
For the remaining two oscillation channels, we have a sizable effect. The oscillation probabilities evaluated
at the one-standard-deviation points for $\theta_{13}$, $-0.17$ and $0.22$, are
\begin{eqnarray}
\mathcal{P}_{e \mu}(\theta_{13}=-0.17)= 0.35,&&\mathcal{P}_{e \mu}(\theta_{13}=0.22)=0.50, \nonumber \\ 
\label{pemex} \\
\mathcal{P}_{\mu \mu}(\theta_{13}=-0.17)= 0.37,&&\mathcal{P}_{\mu \mu}(\theta_{13}=0.22)=0.22. \nonumber \\
\label{pmmex}
\end{eqnarray}
The relative differences are more appropriate quantities to consider; for $\mathcal{P}_{\mu \mu}$ we 
have the most significant effect
\begin{equation}
\frac{\mathcal{P}_{\mu \mu}(\theta_{13}=0.22) - \mathcal{P}_{\mu \mu}(\theta_{13}=-0.17)}
{\mathcal{P}_{\mu \mu}(\theta_{13}=0.22) + \mathcal{P}_{\mu \mu}(\theta_{13}=-0.17)} = -0.25,
\end{equation}
while the effect is still large for $\mathcal{P}_{e \mu}$
\begin{equation}
\frac{\mathcal{P}_{e \mu}(\theta_{13}=0.22) - \mathcal{P}_{e \mu}(\theta_{13}=-0.17)}
{\mathcal{P}_{e \mu}(\theta_{13}=0.22) + \mathcal{P}_{e \mu}(\theta_{13}=-0.17)} =  0.17.
\end{equation}
We compare the one-sigma extremes of $\theta_{13}$ in order to demonstrate the potential size of the
effect.  

\begin{figure}
\includegraphics[width=3in]{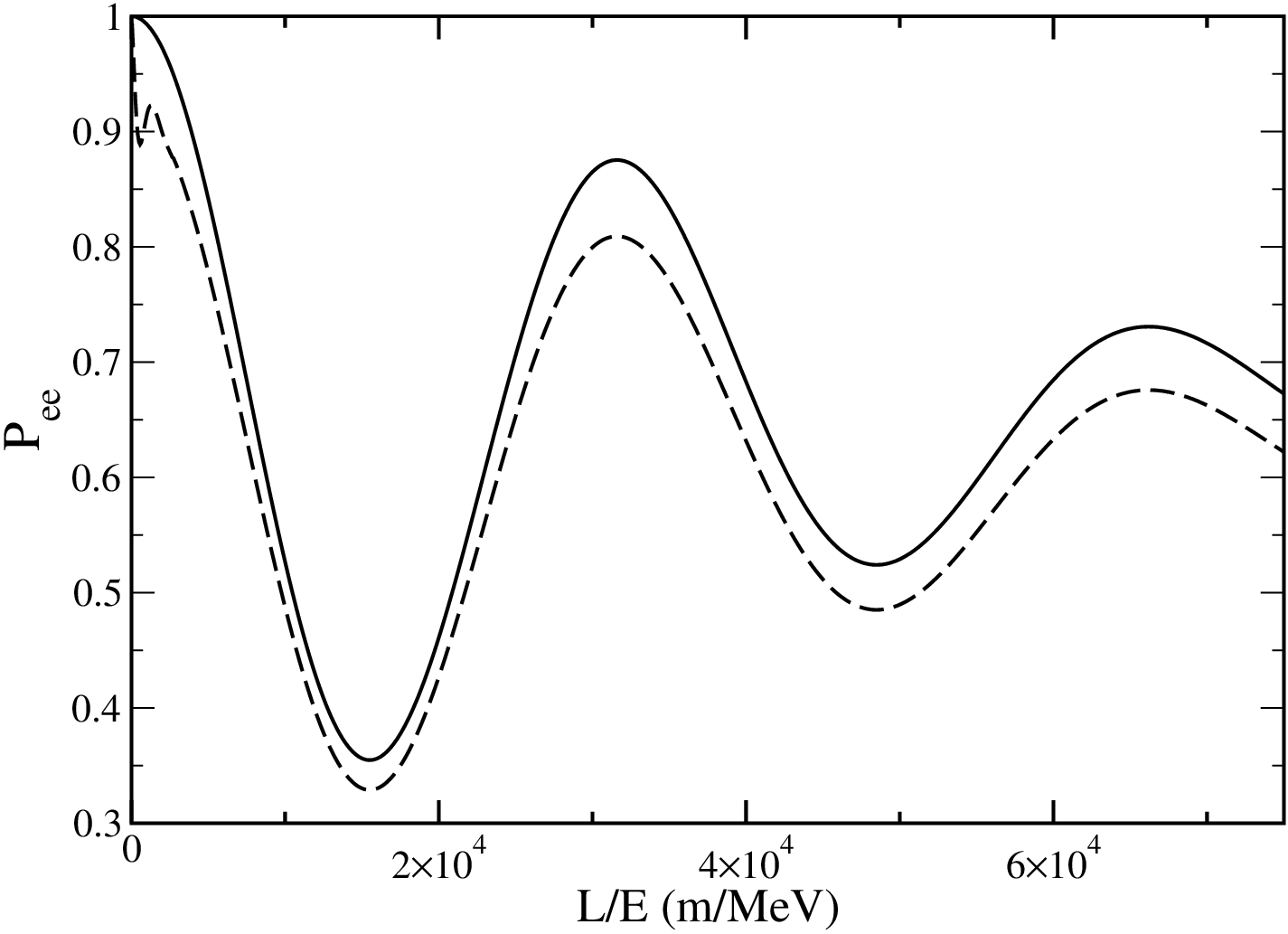}
\caption{${\mathcal P}_{ee}$ versus $L/E$. The solid curve corresponds to the mixing parameters for a fit with $\theta_{13}=0$. 
The dashed curve represents the two curves given by $\theta_{13}=\pm 0.2$ which are identical.
\label{fig2}}
\end{figure}

\section{Simulation}

The previous analytic work tells us where to look if we wish to observe the terms in the oscillation formulae which are linear in $\sin\theta_{13}$. 
We investigate this further by utilizing the analysis from Ref.~\cite{us}. We proceed by fixing all of the oscillation parameters except $\theta_{13}$ to
their values that are given by an analysis  in which $\theta_{13}$ is set to zero. 

\begin{figure}
\includegraphics[width=3in]{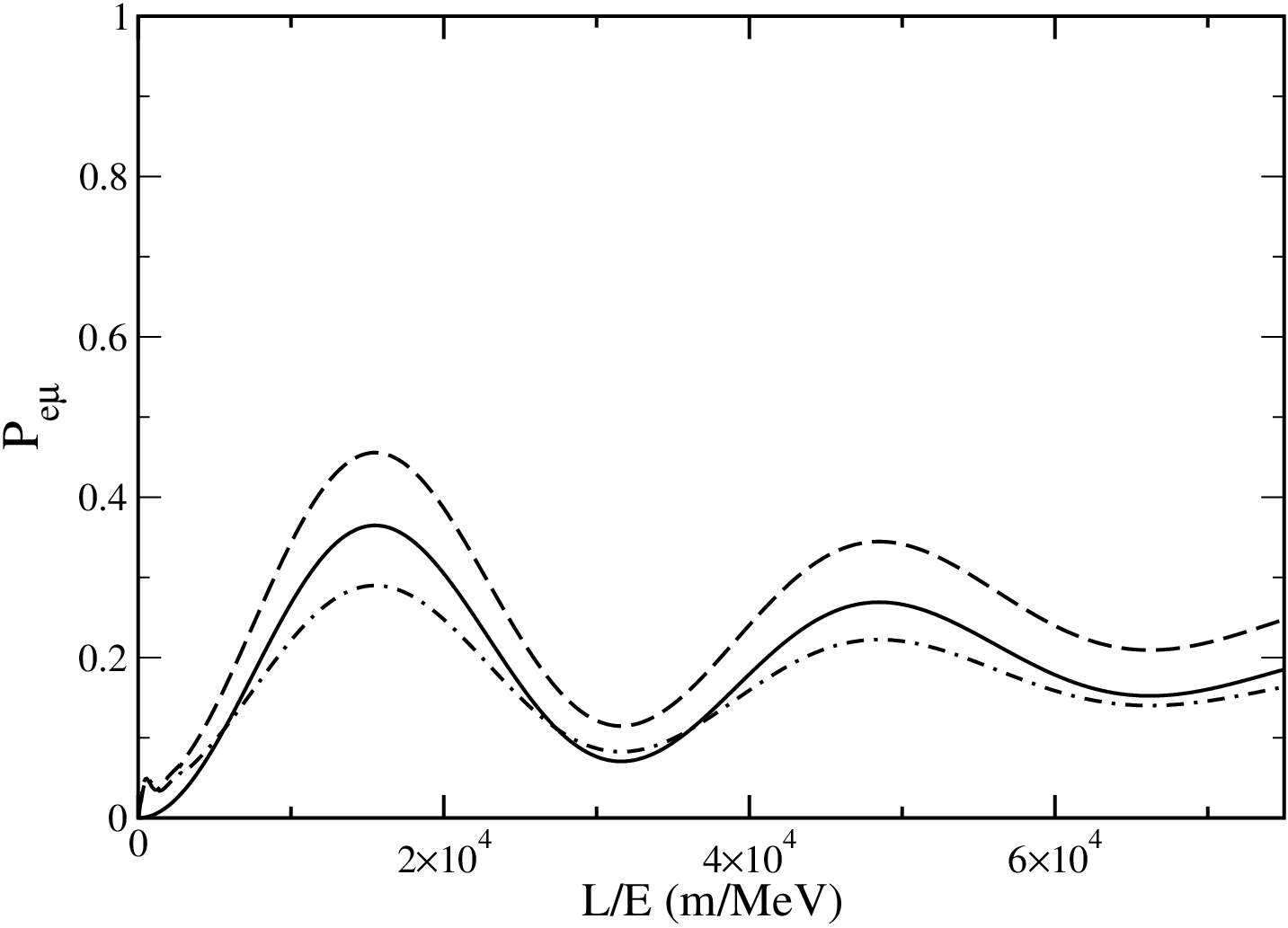}
\caption{${\mathcal P}_{e\mu}$ versus $L/E$. The solid curve corresponds to the mixing parameters for a fit with $\theta_{13}=0$.
The dashed (dot-dashed) curve corresponds to $\theta_{13}=0.2$ ($-0.2$).
\label{fig3}}
\end{figure}

\begin{figure}
\includegraphics[width=3in]{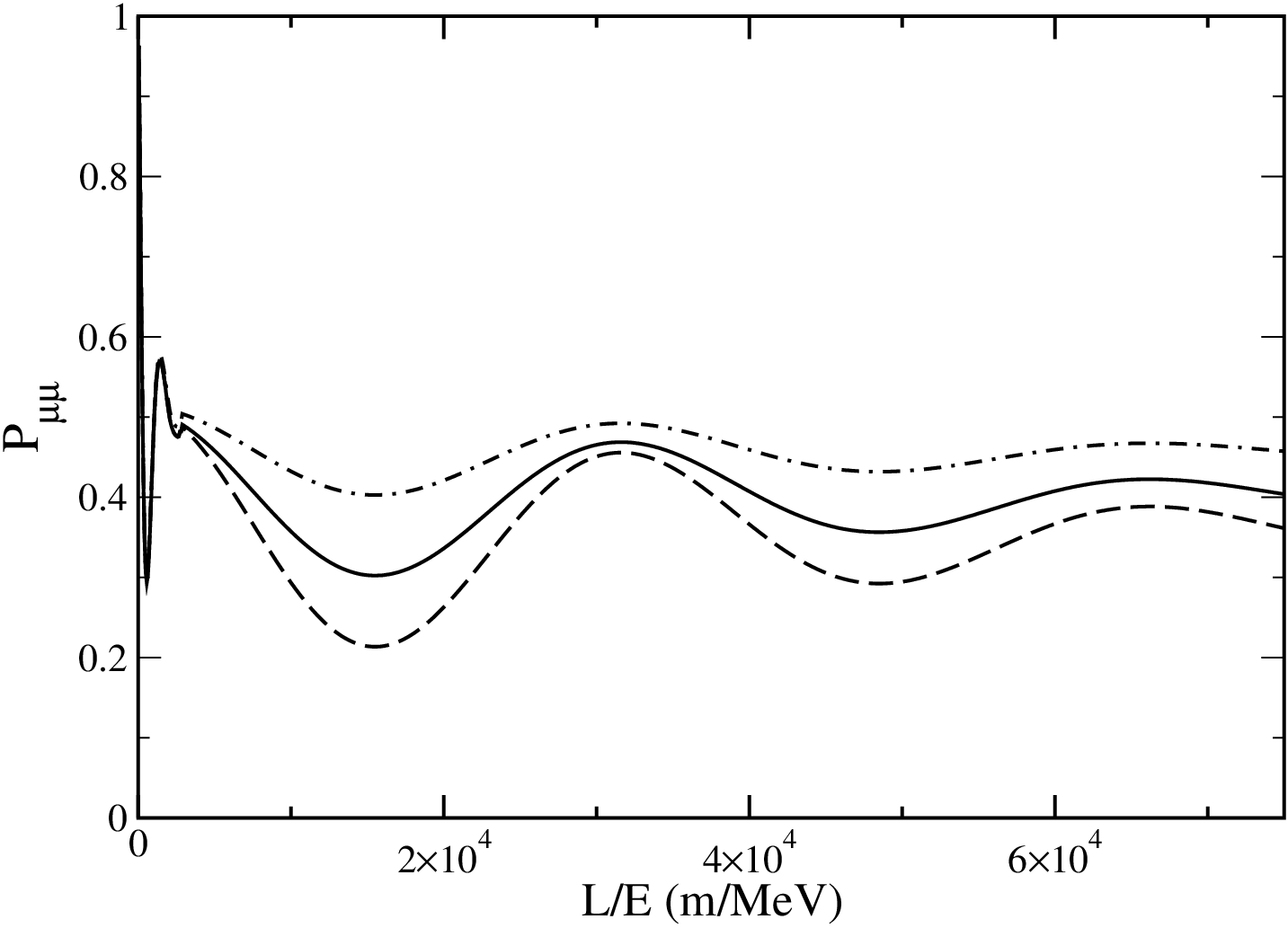}
\caption{The same as Fig.~\protect\ref{fig3} except the muon survival probability ${\mathcal P}_{\mu\mu}$ is presented.
\label{fig4}}
\end{figure}

In Figs.~\ref{fig2}-\ref{fig4}, we present the oscillation probabilities ${\mathcal P}_{ee}$, ${\mathcal P}_{e\mu}$, and ${\mathcal P}_{\mu\mu}$ as
a function of $L/E$  respectively. In all cases we assume a Gaussian spread in energy of twenty percent. In all cases curves are presented for 
$\theta_{13}=0$ and $\theta_{13}=\pm 0.2$. In Fig.~\ref{fig2} for ${\mathcal P}_{ee}$ there are only two curves as ${\mathcal P}_{ee}$ is a function 
of $\sin^2 \theta_{13}$. Thus the curves for $\theta_{13}=\pm 0.2$ are identical. The curve is presented for completeness and to note that there is a
measurable dependence of ${\mathcal P}_{ee}$ on the magnitude of $\theta_{13}$ near the peak at $L/E= 3.2 \times 10^4$ m/MeV, near the location of KamLAND 
as has been indicated in Ref.~\cite{us}. 

In Fig.~\ref{fig3} we verify two facts as derived analytically in the previous section. First, there is a significant linear dependence of 
${\mathcal P}_{e\mu}$ on $\sin \theta_{13}$. Secondly, the dependence is maximal at $\varphi_{12} = \pi /2$ and $3\,\pi /2$. The optimal value of $L/E$
to measure $\theta_{13}$, including its sign, is thus $L/E = 2\pi/\Delta_{21}$. The linear term in $\theta_{13}$ clearly dominates near the maximum
of the oscillation. However, the constant term in Eq.~(\ref{pem}) proportional to $\sin^2 \theta_{13}$ becomes relevant whenever the 
probability is minimal.

In Fig.~\ref{fig4} we see that ${\mathcal P}_{\mu\mu}$ is a somewhat better quantity to measure than ${\mathcal P}_{e\mu}$, as here the linear term is even 
more dominant. Again, the linear dependence on $\theta_{13}$ is maximal at $\varphi_{12} = \pi/2$ and $3\,\pi/2$. 
Fig.~\ref{fig1} demonstrates the inadequacy of parameterizing oscillation parameters as a function of $\sin^2 \theta_{13}$. Figs.~\ref{fig3} and \ref{fig4}
demonstrate this more dramatically as the linear term in $\theta_{13}$ is found to dominate the quadratic term for ${\mathcal P}_{e\mu}$ and
${\mathcal P}_{\mu\mu}$ in this region of $L/E$.

The ideal measurement would be to determine ${\mathcal P}_{\mu\mu}$ over a range for $L/E$ from approximately $1.6\times 10^{3}$ m/MeV to 
$1.6\times 10^{4}$ m/MeV. The lower end of this range provides an overall calibration point where the affect of non-zero $\theta_{13}$ 
is small while the upper end is the point where the effect is largest. To avoid the further complications of the Earth's MSW effect, the energy
should be less than about 100 MeV, the energy of the MSW resonance in the mantle of the Earth. For an upper limit on the energy of 50 MeV, 
this sets an ideal value $L = 80$ km and a range in energy for the muon neutrinos of 5 to 50 MeV. For a smaller value of $L$, the 
range of the neutrino energies would have to be proportionally smaller. 
More likely one's neutrino source would produce neutrinos with energies of at least a few hundred MeV's.  Here, matter effects become significant 
though the qualitative features of the curves in Figs.~(\ref{fig3}) and (\ref{fig4}) can be salvaged.  The most notable changes for oscillations 
through the mantle in this scenario are the increased frequency of oscillations, necessitating a suitable change in the baseline, and the adjustment 
of the matter mixing angle $\theta_{12}^{\, m}$.  However, the difference between the matter mixing angle $\theta_{12}^{\, m}$ and the vacuum value of the mixing angle 
can be minimized by choosing a neutrino energy twice that of the resonance energy.  For such a situation, the qualitative features of the oscillations 
presented above remain the same.

\begin{figure}
\includegraphics[width=3in]{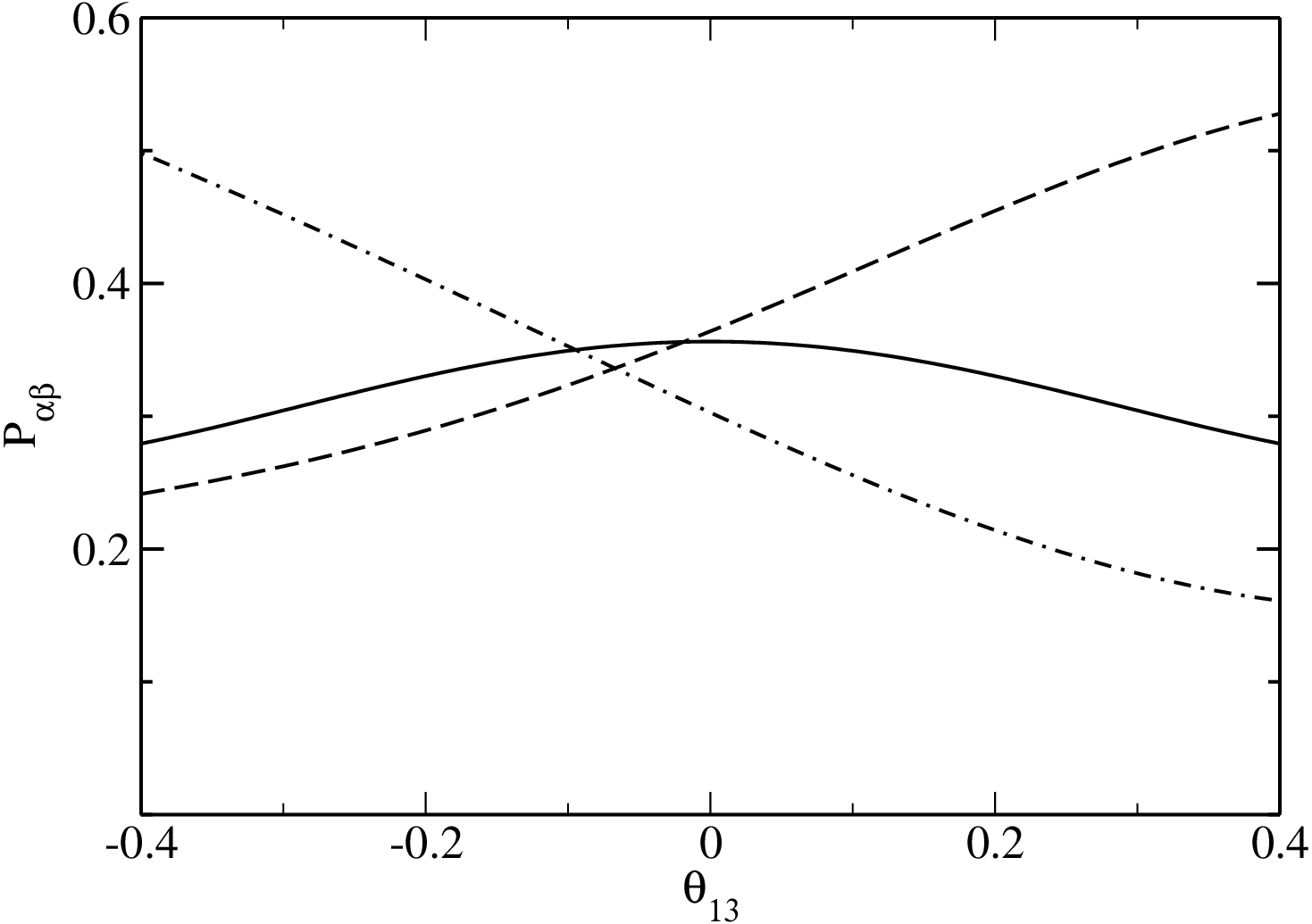}
\caption{The oscillation probabilities versus the mixing angle $\theta_{13}$ with the other parameters fixed at their value from a fit with $\theta_{13}=0$.
The solid curve is ${\mathcal P}_{ee}$, the dashed curve ${\mathcal P}_{e\mu}$, the dot-dashed
curve ${\mathcal P}_{\mu\mu}$. We take $L/E= 1.6\times 10^4$ m/MeV where the effect of the linear terms in $\theta_{13}$ are maximized.
\label{fig5}}
\end{figure}

Finally, in Fig.~\ref{fig5}, we present the values of ${\mathcal P}_{ee}$, ${\mathcal P}_{e\mu}$, and ${\mathcal P}_{\mu\mu}$ as a function of 
$\theta_{13}$. The other parameters are fixed at their values for a fit with $\theta_{13}=0$. We use a value of 
$L/E=1.6\times 10^4$ m/MeV chosen 
so as to maximize the relative importance of the linear terms in $\theta_{13}$. Again, ${\mathcal P}_{ee}$ is quadratic in
$\theta_{13}$. However, ${\mathcal P}_{e\mu}$ and ${\mathcal P}_{\mu\mu}$ are nearly linear in $\theta_{13}$ over this rather large range of $-0.4 \le
\theta_{13}\le 0.4$. The near linearity reinforces our observation that the measurement of 
${\mathcal P}_{e\mu}$ and ${\mathcal P}_{\mu\mu}$ at $L/E = 2\pi/\Delta_{21}$ is a way of determining $\theta_{13}$, including its sign. 

The phenomenology of neutrino oscillations, in the absence of CP violation and various exotica such as a fourth sterile neutrino, has been performed in 
the context of determining three mixing angles and two mass-squared differences. Historically, the results were presented in terms on $\sin^2\theta_{13}$,
which yields an upper limit. When the bounds on the mixing angles were explicitly quoted, they were
all stated to be bounded by $\pi/2$. In Ref.~\cite{gluza}, it was shown that, in the physical case which necessarily includes the MSW matter effect, a 
second branch corresponding to $\delta=\pi$ must also be included. In 
Ref.~\cite{angles}, we extended the derivation, again in the physical case which necessarily includes the MSW matter effect, to show that only the 
$\delta=0$ branch is required if the mixing angle $\theta_{13}$ is allowed to vary from $-\pi/2$ to $+\pi/2$. There are two advantages to this 
convention. First, the allowed region for $\theta_{13}$ consists of a single region that extends on either side of 0, rather than two disjoint regions, 
one for $\delta=0$ and one for $\delta=\pi$.  Secondly, in the presence of CP 
violation, the CP phase is bounded by 0 and $\pi$. Thus a measurement that is sensitive to $\cos\delta$ could uniquely determine the quadrant in which 
$\delta$  lies.

\section{Conclusion}

The natural question that arises from this formal work is whether you can measure $\theta_{13}$ including its sign. In Ref.~\cite{nolsnd3} we demonstrated
that for a model analysis of the world's data, the $\chi^2$ space was asymmetric in $\theta_{13}$, thus demonstrating that such a measurement might be 
possible. A more recent analysis \cite{us} which includes recent data verified this. Here we address the question directly and demonstrate that the 
answer to our question is yes
-- the mixing angle $\theta_{13}$ does appear linearly in the oscillation probabilities at a level where it dominates the quadratic term for the correctly chosen 
oscillation probabilities when measured at the correct value of $L/E$.

We find that this is true for measurements of ${\mathcal P}_{e\mu}$ and ${\mathcal P}_{\mu\mu}$ at $L/E = 2\pi/\Delta_{21}$. This fortuitously also
corresponds to a value of $L/E$ where the contribution from the CP violating effects is minimal. A consequence of this work is 
that it reinforces our earlier thesis that parameterizing neutrino oscillation probabilities as a function of $\sin^2\theta_{13}$ is inadequate. 

There remains additional work to be done. We find that present data indicate a correlation between the allowed value of $\theta_{13}$ and $\theta_{23}$ 
when the linear terms are included in the analysis.
We are investigating the implications for $\theta_{13}$ that will result from measurements of $\theta_{23}$. We have assumed no CP violation in our model 
analysis. Since the CP phase and $\theta_{13}$ are interrelated, this needs further clarification. For cleanliness of interpretation, we propose
do do experiments in a region of $E$ where the Earth MSW effect is small. We are examining in a quantitative way how the Earth MSW effect might modify
an analysis should this be necessary. We, like others, have excluded the LSND 
\cite{LSND} experiments. If MiniBoone verifies the LSND results, then a whole new physics will be needed to
reach a consistent understanding
of neutrino oscillations.

\acknowledgements

Work supported, in part, by the US Department of Energy under contract DE-FG02-96ER40975. The authors thank D.\ V.\ Ahluwalia-Khalilova and S.\ 
Palomares-Ruiz for helpful conversations.

\bibliography{th13expc2}

\begin{thebibliography}{1}

\bibitem{LSND}
C. Athanassopoulos, et al., Phys. Rev. Lett. {\bf 77}, 3082 (1996); Phys. Rev. C {\bf 54},
2685 (1996); Phys. Rev. Lett. {\bf 81},1774 (1998); Phys. Rev. C {\bf 58}, (2489); A. 
Aguilar et al., Phys. Rev. D {\bf 64}, 112007 (2001).

\bibitem{bargerrev}
V. Barger, D. Marfatia, and K. Whisnant, Int. J. Mod. Phys. E {\bf 12}, 569 (2003).

\bibitem{k2k}
M.~H.~Ahn {\it et al.}  [K2K Collaboration],
Phys. Rev. Lett. {\bf 90}, 041801 (2003);
{\bf 93}, 051801 (2004).

\bibitem{superk}
Y. Fukuda et al., Phys. Lett. {\bf B436}, 33 (1998); Phys. Rev. Lett. {\bf 82}, 2644 
(1999); {\bf 86}, 5651 (2001).

\bibitem{minos}
R.~Saakian  [MINOS Collaboration],
Phys. Atom. Nucl.  {\bf 67}, 1084 (2004)
[Yad. Fiz.  {\bf 67}, 1112 (2004)].

\bibitem{opera}
M.~Dracos  [OPERA Collaboration],
Phys. Atom. Nucl.  {\bf 67}, 1092 (2004)
[Yad. Fiz.  {\bf 67}, 1120 (2004)].

\bibitem{icarus}
J.~Lagoda  [ICARUS Collaboration],
Phys. Atom. Nucl.  {\bf 67}, 1107 (2004)
[Yad. Fiz.  {\bf 67}, 1135 (2004)].

\bibitem{lblanalysis}
V.~D.~Barger, A.~M.~Gago, D.~Marfatia, W.~J.~C.~Teves, B.~P.~Wood and R.~Zukanovich Funchal,
Phys. Rev. D {\bf 65}, 053016 (2002).

\bibitem{eightfold}
V.~Barger, D.~Marfatia and K.~Whisnant, Phys. Rev. D {\bf 65}, 073023 (2002).

\bibitem{angles}
D.~C.~Latimer and D.~J.~Ernst, Phys. Rev. D {\bf 71}, 017301 (2005).

\bibitem{msw}
L. Wolfenstein, Phys. Rev. D {\bf 17}, 2369 (1978); 
S. P. Mikheyev and A. Yu. Smirnov, Sov. J. Nucl. Phys. {\bf 42}, 913 (1985).

\bibitem{nolsnd3}
D. C. Latimer and D. J. Ernst, nucl-th/0404059.

\bibitem{us}
D. C. Latimer and D. J. Ernst, in preparation.

\bibitem{smirnov1}
O. L. G. Peres and A. Yu. Smirnov, Nucl. Phys. B {\bf 680}, 479
(2004).

\bibitem{eexcess}
G. L. Fogli, E. Lisi, A. Marrone, and G. Scioscia, Phys. Rev. D {\bf
59}, 033001 (1998);
C. W. Kim and U. W. Lee, Phys. Lett. {\bf B444}, 204 (1998);
O. L. G. Peres and A. Yu. Smirnov, Phys. Lett. {\bf B456}, 204
(1999);
O. L. G. Peres and A. Yu. Smirnov, Nucl. Phys. B (Proc. Suppl.) 
{\bf 110}, 355 (2002).

\bibitem{expans}
E. K. Akhmedov, R. Johansson, M. Lindner, T. Ohlsson, and T. Schwetz,
JHEP {\bf 0404}, 078 (2004).

\bibitem{PDG}
Particle Data Group, Phys. Lett. {\bf B592} (2004)

\bibitem{solar}
B. T. Cleveland, et al., Astrophys. J. {\bf 496}, 505 (1998);
J. N. Abdurashitov, et al., Phys. Rev. C {\bf 60}, 055801 (1999);
J. Exp. Theor. Phys. {\bf 95}, 181 (2002);
W. Hampel, et al., Phys. Lett. {\bf B447}, 127 (1999);
M. Altmann, et al., Phys. Lett. {\bf B490}, 16 (2000);
Q. R. Ahmad, et al., Phys. Rev. Lett. {\bf 87}, 071301 (2001);
Phys. Rev. Lett. {\bf 89}, 011301 (2002);
S. N. Ahmed, et al., nucl-ex/0309004.

\bibitem{reactors}
M. Apollonio, et al., Phys. Lett. {\bf B 420}, 397 (1998); {\bf B466}, 
415 (1999); Eur. Phys. J. {\bf C 27}, 331 (2003);
K. Eguchi, et al., Phys. Rev. Lett. {\bf 90}, 021802 (2003);
T. Araki et al., hep-ex/0406035.

\bibitem{atmo}
Y. Fukuda, et al., Phys. Lett. {\bf B335}, 237 (1994); {\bf B433}, 9 (1998);
{\bf B436}, 33 (1998); Phys. Rev. Lett. {\bf 81}, 1562 (1998);
Phys. Lett. {\bf B436}, 33 (1998); Phys. Rev. Lett. {\bf 82}, 2644 
(1999); {\bf 86}, 5651 (2001); Y.Ashie, et al., Phys. Rev. Lett. {\bf 93}, 
101801 (2004).

\bibitem{gluza}
J. Gluza and M. Zralek, Phys. Lett. {\bf B517}, 158 (2001).

\end{thebibliography}


\end{document}